\documentclass[twocolumn,10pt,superscriptaddress,reprint]{revtex4-1}
\usepackage{xcolor}
\usepackage{graphicx}
\usepackage{amsmath}
\begin{document}

\title{Particle velocity controls phase transitions in contagion dynamics}

\author{Jorge P. Rodr\'iguez}
\address{Instituto de F\'isica Interdisciplinar y Sistemas Complejos IFISC (CSIC-UIB), Palma de Mallorca, E-07122, Spain}
\email{jorgeprodriguezg (at) gmail (dot) com}
\author{Fakhteh Ghanbarnejad}
\address{Technische Universit\"at Berlin, Berlin, 10623, Germany}
\address{The Abdus Salam International Centre for Theoretical Physics (ICTP), Trieste, 34151, Italy}
\email{fakhteh.ghanbarnejad (at) gmail (dot) com}
\author{V\'ictor M. Egu\'iluz}
\address{Instituto de F\'isica Interdisciplinar y Sistemas Complejos IFISC (CSIC-UIB), Palma de Mallorca, E-07122, Spain}

\begin{abstract}
Interactions often require the proximity between particles. The movement of particles, thus, drives the change of the neighbors which are located in their proximity, leading to a sequence of interactions. In pathogenic contagion, infections occur through proximal interactions, but at the same time the movement facilitates the co-location of different strains. We analyze how the particle velocity impacts on the phase transitions on the contagion process of both a single infection and two cooperative infections. First, we identify an optimal velocity (close to half of the interaction range normalized by the recovery time) associated with the largest epidemic threshold, such that decreasing the velocity below the optimal value leads to larger outbreaks. Second, in the cooperative case, the system displays a continuous transition for low velocities, which becomes discontinuous for velocities of the order of three times the optimal velocity. Finally, we describe these characteristic regimes and explain the mechanisms driving the dynamics.
\end{abstract}

\flushbottom
\maketitle

\thispagestyle{empty}

\section*{Introduction}

Spreading processes, such as contagion or rumor transmission, have been modeled using agent-based approaches \cite{pastor2001epidemic,nekovee2007theory}. These dynamics reach a higher level of complexity when, apart from the diffusive nature of the process, the particles move. For example, some oviparous fauna release sperm and eggs in aquatic environments; thus, the fertilization events happen via mobility-driven encounters of these cells \cite{de2002monopolization}. Mobility impacts biological contagion, such as the dispersal of fungi spores in plant fields \cite{brown2002aerial}, or the appearance of microcolonies through the attachment to the particles being carried inside the xylem for plants \cite{yadeta2013xylem} or the veins for animals \cite{ribet2015bacterial}. Pathogens do not spread only inside a single individual body, rather amongst different individuals too, such that proximity allows the contagion from one organism to another; this has led to a diversity of studies of epidemiology in systems of mobile fauna introducing, for example, the appearance of large outbreaks when a pathogen switches from spreading in one species to another \cite{parrish2008cross}, or the particularities of waterborne infections in marine environments, such as fishing or death-based transmission \cite{bidegain2016marine,laferty2017marine}. Indeed, aquaculture represents a key industry for nutritional and financial security in developing countries, and can be strongly affected by infections spreading among the confined but mobile fishes \cite{leung2013more,stentiford2017new}. Additionally, the coupling between movement and spreading dynamics is also relevant in socio-technological systems, where malware is transmitted through the connection of multiple devices to WiFi and Bluetooth networks, spreading further because as time evolves the devices are connected to different networks due to the mobility of their users \cite{hu2009wifi}. Hence, mobility drives the appearance of temporal connections, which modify the statistical properties of the spreading processes \cite{holme2012temporal,masuda2013predicting}.

Hosts' movement facilitated the spread of infections that, throughout the human history, have affected large populations \cite{taubenberger20061918,johnson2002updating,potter2001history}. In fact, several infections that emerged in geographically distant locations may converge to the same environment due to their hosts' mobility, facilitating the interaction between different pathogens \cite{alvar1997leishmania,elberg1957cross}, either cooperating or competing \cite{abu2008interactions,sanz2014dynamics,wang2019coevolution}. These interactions modify the infection rates, such that the infection rate of a disease depends on the history of other infections of the host. This implies that the primary infection rate, which is the infection rate for a host that has never been infected with any disease, is different from the secondary one, which represents the infection rate for agents that have been previously infected with one disease; and these will be different from the tertiary, quaternary, and higher order infection rates. For example, competition between two different infections can lead to cross-immunity, that is, the particles are immunized against other infections after a primary infection \cite{newman2005threshold,karrer2011competing} (\emph{i.e.}, the secondary infection rates are zero). In contrast, cooperation represents the weakening of the individuals after a primary infection, increasing the secondary infection rates \cite{chen2013outbreaks,cai2015avalanche,chen2017fundamental,rodriguez2018diversity,zarei2019exact}.

Spreading dynamics coupled with particle movement has been typically modelled with metapopulation approaches, which consider the geographical locations as patches of particles connected through fluxes (which define an adjacency matrix). For example, a recent work showed the presence of a region in parameter space where the mobility has a detrimental effect on the outbreak size \cite{gomez2018critical}. However, metapopulation approaches are limited for representing the temporal aspects of the connecting fluxes and the structured interaction inside each patch. Hence, agent-based approaches, where the particles interact with their proximal neighbors, are useful not only for determining the microscopic mechanisms leading to large outbreaks, but also which movement patterns hinder the spreading process. Our approach is based on a temporal network given by proximal interactions, determined by the distance between mobile particles, in the spirit of temporal contact networks, which consist on empirical measurements of who interacts with whom and when. For example, contact tracing has been used for analyzing the structure of the social structure underlying the spread of sexually transmitted diseases \cite{wylie2001patterns}. In fact, correlations and the circadian activity patterns lead to discontinuous transitions in the spreading of two cooperative infections \cite{rodriguez2017risk}. For single infection dynamics on empirical contact networks, the ratio between the time scales of network dynamics and epidemic dynamics influences the epidemic threshold, such that when the network dynamics is faster, the epidemic threshold is smaller \cite{speidel2016temporal}.

In a scenario where particles move, we study the contagion dynamics of either a single or two cooperative infections arguing about the role of particle velocity as a control parameter. First, we introduce the model describing the contagion process and the particle movement. Second, we analyze the behavior of a single infection spreading among a population of mobile particles. Finally, we extend our analysis to the dynamics of two cooperative infections.

\section*{Model}

As initial condition, we distribute $N$ particles randomly in a two-dimensional space of size $L\times L$ with periodic boundary conditions. Each particle $i$ is assigned a direction of movement $\xi_i \in [0,2\pi)$ randomly from a uniform distribution. This direction of movement does not change with time. The movement is uniform rectilinear with velocity $v$, such that the time evolution of the coordinates of particle $i$, $x_i(t)$ and $y_i(t)$, is given by
\begin{equation}
\begin{array}{l}
x_i(t+1)=\left[x_i(t)+v\cos \xi_i \right] \bmod L \\
y_i(t+1)=\left[y_i(t)+v\sin \xi_i \right] \bmod L
\end{array}
\label{motion}
\end{equation} 
where the motion direction is represented by the angles $\xi_i$ and $v$ is the traveled distance per unit of time. The  Susceptible-Infected-Recovered (SIR) model describes the dynamics \cite{kermack1927contribution} between three states: an infected particle transmits the infection with probability $p$ to each particle located at a distance smaller than $d$, and recovers after one time step. The time scale of the dynamics is set by the time it takes one infected particle to recover, which we define as a time unit. For two cooperative infections, $\mathcal{A}$ and $\mathcal{B}$, the dynamics is similar to the SIR model \cite{chen2013outbreaks}, with primary infections from an infected neighbor happening with probability $p$, while secondary infections occur with probability $q$.

\begin{figure*}[bt!]
\centering
\includegraphics[width=0.8\textwidth]{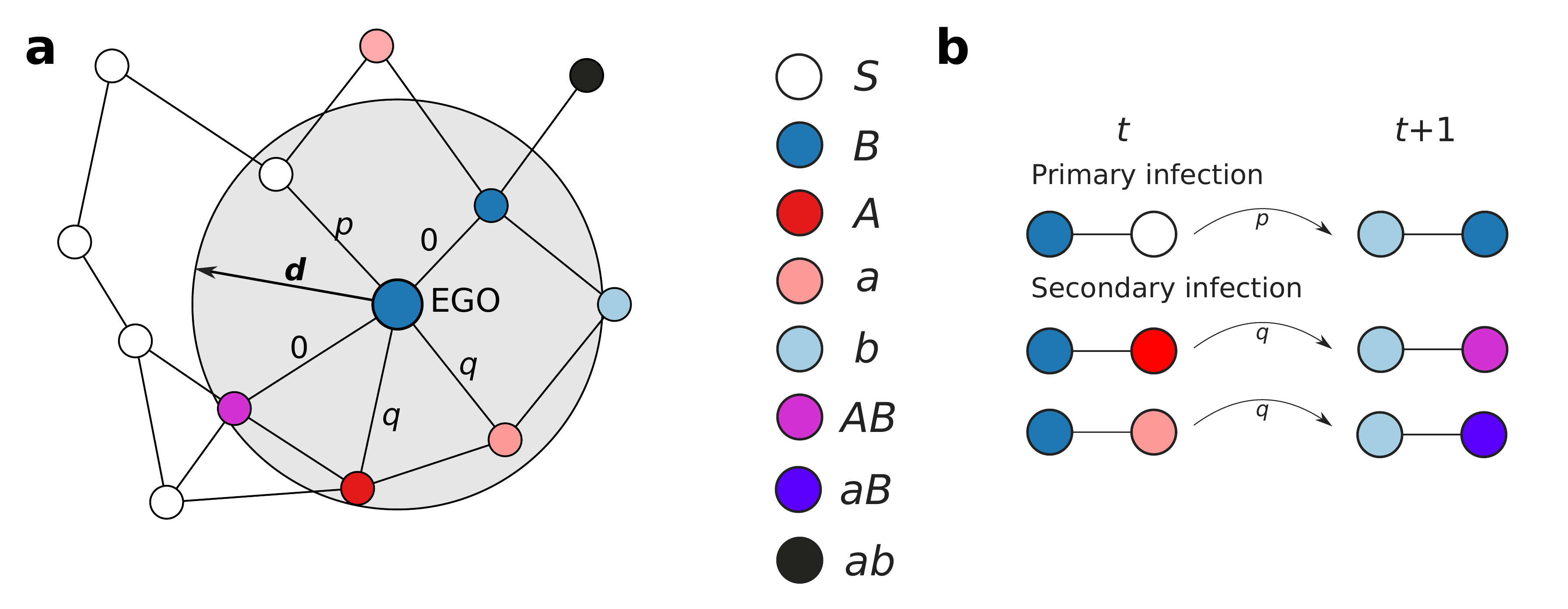}
\caption{Scheme of the model. {\bf a}, Contagion of infection $\mathcal{B}$ from EGO to its neighbors (particles at a distance smaller than $d$). Labels on the links indicate the infection probabilities. {\bf b}, Dynamics of primary and secondary infections, happening with probabilities $p$ and $q$, respectively.}
\label{scheme}
\end{figure*} 

For a single infection, the initially infected population does not transmit the infection for probabilities below a critical value, $p<p_c$. However, for $p>p_c$, the infection affects a finite fraction of the system \cite{pastor2015epidemic}. For two cooperative infections, the case $q=p$ represents two independent contagion processes, while for $q>p$ both infections have a cooperative interaction. Particularly, focusing on an infected particle EGO, each neighbor particle $i$ updates its state according to the following rules (Fig. \ref{scheme}): a) if $i$ is in state $S$, it will be primarily infected with probability $p$; b) if $i$ is infected or recovered with/from the same infection as EGO, nothing will happen; c) if $i$ is infected/recovered with/from the other infection, it will be secondarily infected with probability $q$. In our simulations, we initially assign the state $AB$ ($I$ for single infection) to a randomly chosen particle and the state $S$ to the rest. At each time step, first the state of all the particles is updated (Fig. \ref{scheme}); and then, their positions are updated synchronously according to Eq. \ref{motion}.

At each time step, the set of interactions is described as a random geometric graph \cite{penrose2003random}, because the initial positions and the movement directions are uniformly distributed. In that topology, the expected number of neighbors is 

\begin{equation}
\langle k \rangle=(N-1)\frac{\pi d^2}{L^2}
\label{avdeg}
\end{equation}
We identify two referential spatial scales in random geometric graphs. First, random geometric graphs with periodic boundary conditions display a percolation transition \cite{dall2002random} in 2D at $\langle k \rangle_c\approx 4.52$. Thus, fixing $N=2^{12}$ and $L=1280$ (same particle density as in Ref. \cite{fujiwara2011synchronization}), we obtain $d_c=L\sqrt{\frac{\langle k \rangle_c}{(N-1)\pi}}\approx 24$. Second, the system is fully connected for $d \ge d_{\text{max}}=\frac{L}{\sqrt{2}}\approx 905$.

The temporal network describing the contacts in our system displays temporal correlations. In fact, temporal correlations are related to the probability that, if there is a contact between particles at time $t$, this contact remains at time $t+1$. Considering the particle velocity as a control parameter, the temporal correlations have two limits. One limit is the static case, where the particles do not move, such that the set of contacts does not change in time (Fig. S1), and there will be finite outbreaks only if $d>d_c$ (necessary but not sufficient condition). In the other limit, particles move so fast such that the set of contacts between two consecutive updates are uncorrelated, and can be represented by generating a random location for each particle at every update. Thus, the probability to observe the same interaction or new interactions in two consecutive time steps can be used to link the velocity (one of our control parameters) to empirical temporal networks, such as those specifying human contacts measured through radio frequency identification devices \cite{vanhems2013estimating,barrat2013empirical}.

Recapitulating, our model has three control parameters ($d$, $v$, $p$). We will keep one of these parameters constant for studying the behavior of the order parameter, which will be the density of (doubly) recovered particles $\rho$ ($\rho_{\text{ab}}$) in the final absorbing configuration for single (two cooperative) infection(s).

\section*{Results}
\subsection*{Single infection}

\begin{figure*}[bt!]
\centering
\includegraphics[width=0.8\textwidth]{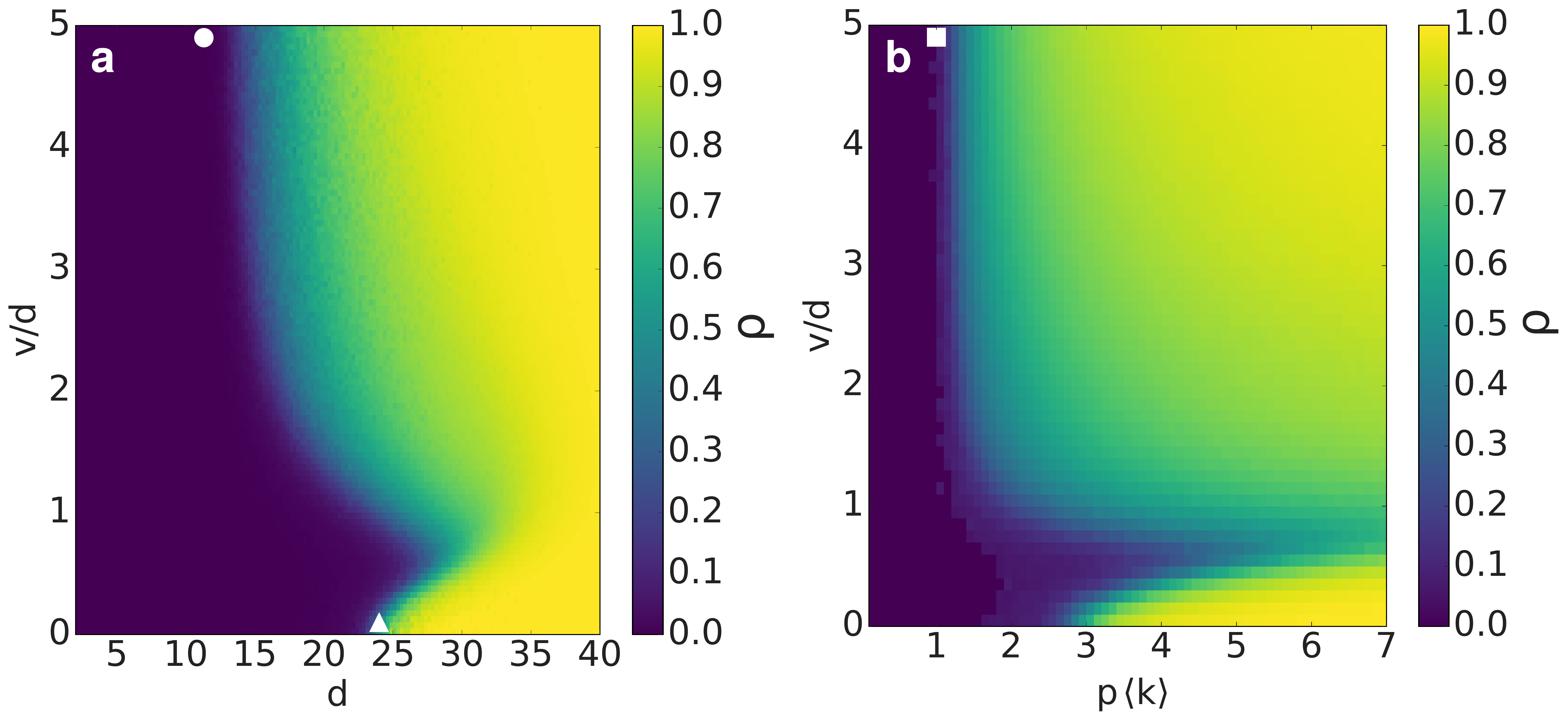}
\caption{Single infection: fraction $\rho$ of recovered particles in the final absorbing configuration for: {\bf a}, maximum infection probability ($p=1$) in the $(d,v)$ space; {\bf b}, $d=30\sim d_c$, in the $(p,v)$ space. Symbols depict the cases a, $d_{\text{min}}$ (circle) and $d_c$ (triangle); b, $p\langle k \rangle=1$ (square).}
\label{onedisvaryv}
\end{figure*}

We start studying the case of maximum infection probability ($p=1$) in the $(d,v)$ space. For a given particle velocity $v$, there is a threshold interaction range $d^*(v)$ such that the infection spreads to a finite fraction of the population for $d>d^*$ (Fig. \ref{onedisvaryv}a). For very high particle velocities, this threshold approaches an asymptotic value $d_{min}$ (circle in Fig. \ref{onedisvaryv}a) while, as the velocity is decreased, $d^*$ increases, reaching a maximum for $v\approx 0.5d$ and then decreasing to $d^*(v=0)=d_c$ (triangle in Fig. \ref{onedisvaryv}a). Thus, we identify four different regimes when, for a fixed $d$, the particle velocity is varied: i) $d<d_{\text{min}}$: the infection does not spread, independently of the velocity;  ii) $d_{\text{min}}<d<d_c$: there is a threshold velocity $v_c$, such that for $v>v_c$ the infection affects a finite fraction of the population; iii) $d\gtrsim d_c$: the infection spreads in the static case ($v=0$), but the low velocities are detrimental for the contagion process, implying that increasing $v$ leads to smaller $\rho$. In this regime, for a given $d$, there are two critical velocities, $v_c^-$ and $v_c^+$, such that there are no macroscopic outbreaks for $v_c^-<v<v_c^+$; iv) $d\gg d_c$: the outbreak reaches all the system, independently of $v$.

In fact, there are two cases that help us understand the limit cases of $d^*$:
\begin{itemize}
\item In the static limit ($v=0$), particles do not move. Hence, the infection spreads along a random geometric graph and, considering $p=1$, it affects a finite fraction of the population only for interaction ranges in the supercritical percolation regime, $d>d_c$ (triangle in Fig. \ref{onedisvaryv}a).
\item The uncorrelated contact sequence limit ($v\rightarrow \infty$) represents a sequence of spatial configurations in which the positions of the particles are randomly reassigned after each time step. This implies that, for a high enough number of particles and $\langle k \rangle \ll N$, there are no common contacts in two consecutive time steps. For an infection probability $p$, the dynamics in this limit, known as annealed network, is described by the mean-field approximation for SIR dynamics \cite{pastor2015epidemic}, where $\rho(p)$ follows the implicit equation $\rho=1-e^{-p\langle k \rangle \rho}$. This equation shows the dependence of $\rho$ on $p\langle k \rangle$, and has a non-trivial solution for $p\langle k \rangle >1$. Hence, the epidemic threshold is $p_c=\frac{1}{\langle k \rangle}$. In fact, considering this limit for different interaction ranges, $\rho(p)$ has a universal behavior for all $d$ when $p$ is rescaled by $\frac{1}{\langle k \rangle}$ (Fig. S2). From the mean-field solution, and considering that $p$ is a probability ($p\leq 1$), we deduce that there is a unique solution (the trivial solution, $\rho=0$) for $\langle k \rangle<\langle k \rangle_{\text{min}}=1$. Thus, we estimate $d_{\text{min}}$ substituting $\langle k \rangle_{\text{min}}$ in Eq. \ref{avdeg}, where we obtain $d_{\text{min}}=L\sqrt{\frac{1}{(N-1)\pi}}\approx 11$ (circle in Fig. \ref{onedisvaryv}a).
\end{itemize}

\subsubsection*{Finite particle velocity: non-monotonic behavior}

\begin{figure*}[bt!]
\centering
\includegraphics[width=0.8\textwidth]{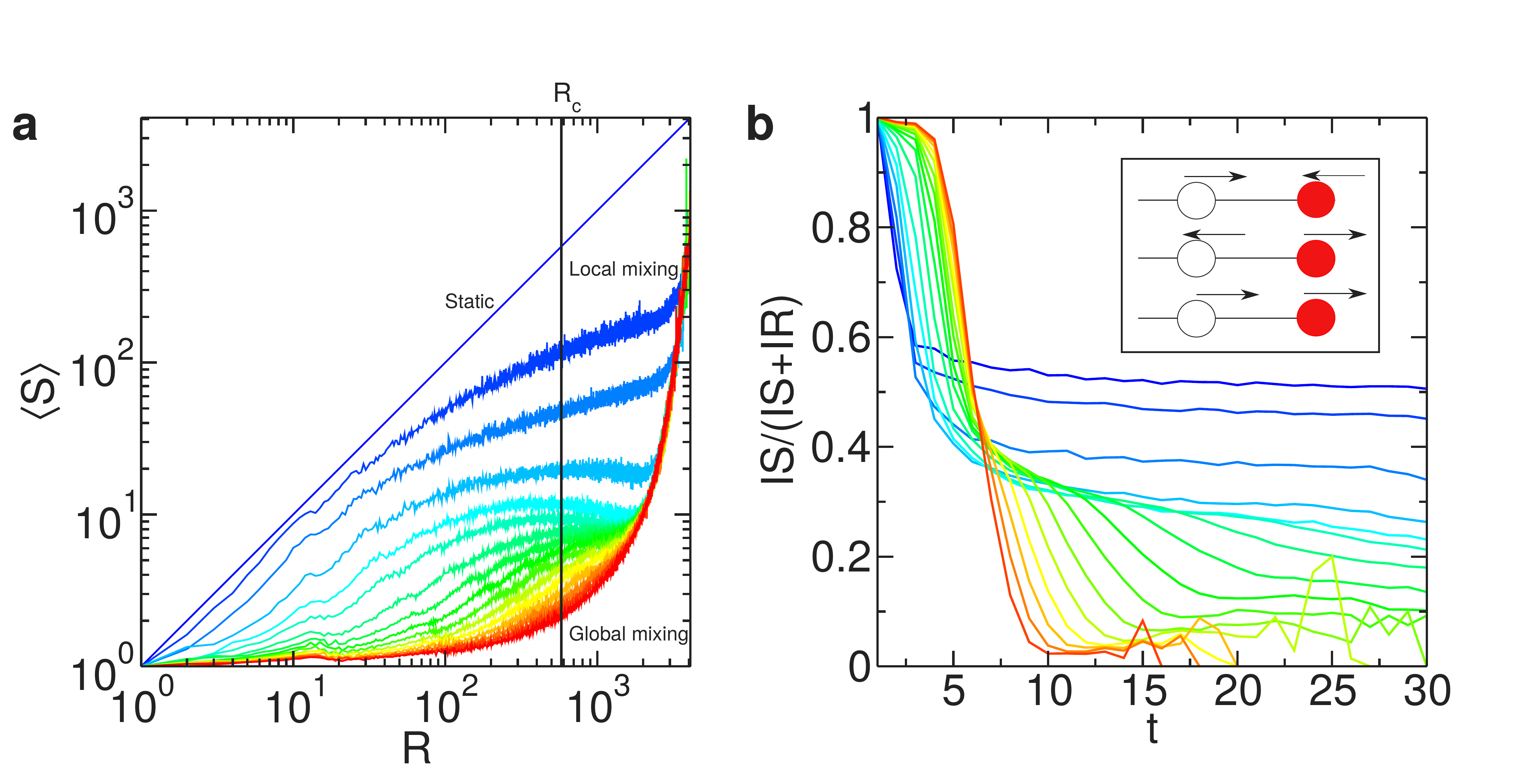}
\caption{Single infection: mechanism for a non-monotonic outbreak size with particle velocity. {\bf a}, Average connected component size $\langle C \rangle_R$ of the network composed by the recovered particles and the contacts between them, as a function of the number of recovered particles $R$. The black vertical line represents the critical value $R_c$ for percolation in a random geometric graph. {\bf b}, Time evolution of the ratio between the number of IS links and the sum of IS and IR links. Inset: the three possible scenarios for a mobile system in 1D, with susceptible and infected particles depicted, respectively, in white and red colors. Velocity is changed from $v=0$ (blue) to $v=d$ (cyan) with $\Delta v=0.2d$, and from $v=d$ to $v=5d$ (red) with $\Delta v=0.4d$, for $d=30\gtrsim d_c$ and $p=1$.}
\label{mechveloc}
\end{figure*}

Previously, we have described a set of interaction ranges ($d\gtrsim d_c$) for which there is a non-monotonic behavior of the outbreak size with the velocity for $p=1$ (Fig. \ref{onedisvaryv}a). Now, we want to study the behavior with velocity for $0\leq p \leq 1$ and for a fixed interaction range $d=30\gtrsim d_c$. Considering the uncorrelated contact sequence limit as a point of reference, where our order parameter $\rho$ depended only on $p\langle k \rangle$, we normalise the infection probability with the average degree $\langle k \rangle (d=30)\approx 7$, such that the case $d=30$ in Fig. \ref{onedisvaryv}a ($p=1$) corresponds to the case $p\langle k \rangle=7$ in Fig. \ref{onedisvaryv}b. We observe that, for a given velocity, there is an epidemic threshold $p_c(v)$, that is, there are macroscopic outbreaks for $p>p_c$, and the epidemic threshold has also a non-monotonic behavior with $v$, approaching the limit $p_c\langle k \rangle=1$ as $v\rightarrow \infty$ (Fig. \ref{onedisvaryv}b). 

For simplicity, we explain the observed non-monotonic behavior for the case $p=1$ (Fig. \ref{onedisvaryv}a). For $v=0$, the infection spreads following an expanding front (Fig. S3); however, as the velocity is increased, the infected particles are further from those that infected them. such that for high velocities there is not a well-defined expanding front and the dynamics is described by a mean-field time evolution (global mixing). We quantify the presence or absence of an expanding front measuring the average connected component size $\langle C \rangle_R$ of recovered particles at each time step (\emph{i.e.}, considering only the recovered particles and the contacts between them) as a function of the number of recovered particles $R$. If the dynamics is driven by an expanding front, $\langle C \rangle_R\sim R$, while if the spreading is driven by global mixing $\langle C \rangle_R \sim R^0$ for low $R$, and for higher values of $R$ there is a percolation transition at $R_c=1+\frac{L^2}{\pi d^2}$ (Fig. \ref{mechveloc}a). Hence, as $\langle C \rangle_{R,v=0}=R$, the contagion among the population is driven by an expanding front in the static case. Moreover, the spatial configurations for $v=0$ and $p=1$ are ordered, as SIR dynamics belongs to the universality class of directed percolation, implying that there are not any contacts between recovered and susceptible particles, and then the system is composed by coherent regions of recovered (the core of the expanding front), infected (the border of the front) and susceptible particles (those out of the front). Hence, the contagion process in the static case represents an expanding front with ordered contacts.

However, for low velocities, the contagion among the population is still described by an expanding front (Fig. \ref{mechveloc}a), but the movement of the particles leads to local mixing at the edge of the front. For example, there are contacts between recovered and susceptible particles, which did not occur in the static case. As the infection is transmitted through the contacts between infected and susceptible particles (IS links), we compare its abundance with the number of contacts between infected and recovered particles (IR links). After an initial transient, the number of IS and the number of IR links are balanced in the static case. However, when velocity is slightly increased, the IS links are less frequent (Fig. \ref{mechveloc}b). Considering that the average degree is fixed by Eq. \ref{avdeg}, this change in the relative abundance implies that each infected particle will have a smaller amount of contacts with susceptible particles. Indeed, as $d\gtrsim d_c$, this decrease represents a dynamical behavior similar to a lower $\langle k \rangle$ (\emph{i.e.}, $d<d_c$), such that the observed discontinuity may represent a dynamical percolation transition.

The decrease in the relative abundance of links between infected and susceptible particles (Fig. S4) is illustrated with a similar set-up in a 1D configuration. In this case, at the edge of an expanding front, there are three possibilities (Fig. \ref{mechveloc}b inset): i) the susceptible and the infected particles are approaching: the susceptible particle will get infected, but at next time step it will be surrounded by many recovered particles; ii) they are getting away: if initially they are separated by $d_0$, at the next time step their distance will be $d_1=d_0+2v$ such that, if $v=0.5d$, $d_1>d$ (optimal velocity); iii) they move in the same direction, leading to the same microscopic behavior as in the case of static particles. In conclusion, this non-monotonic behavior is explained through the cases i) and ii) that, for an expanding front, imply a detrimental effect of low velocities on the contagion.

Finally, for high velocities, the system approaches asymptotically the uncorrelated contact sequence limit. This means that an infected particle at time $t$ is not in contact with the particle that infected it (at time $t-1$), such that the dynamics cannot be described by an expanding front. As a consequence, the initial transient lasts longer (high abundance of IS links, Fig. \ref{mechveloc}b) and, as $p=1$, most of the particles in the system are infected during this transient, leading to $\rho\approx 1$ in the final absorbing configuration.

\subsection*{Two cooperative infections}

\begin{figure}[h!]
\centering
\includegraphics[width=0.5\textwidth]{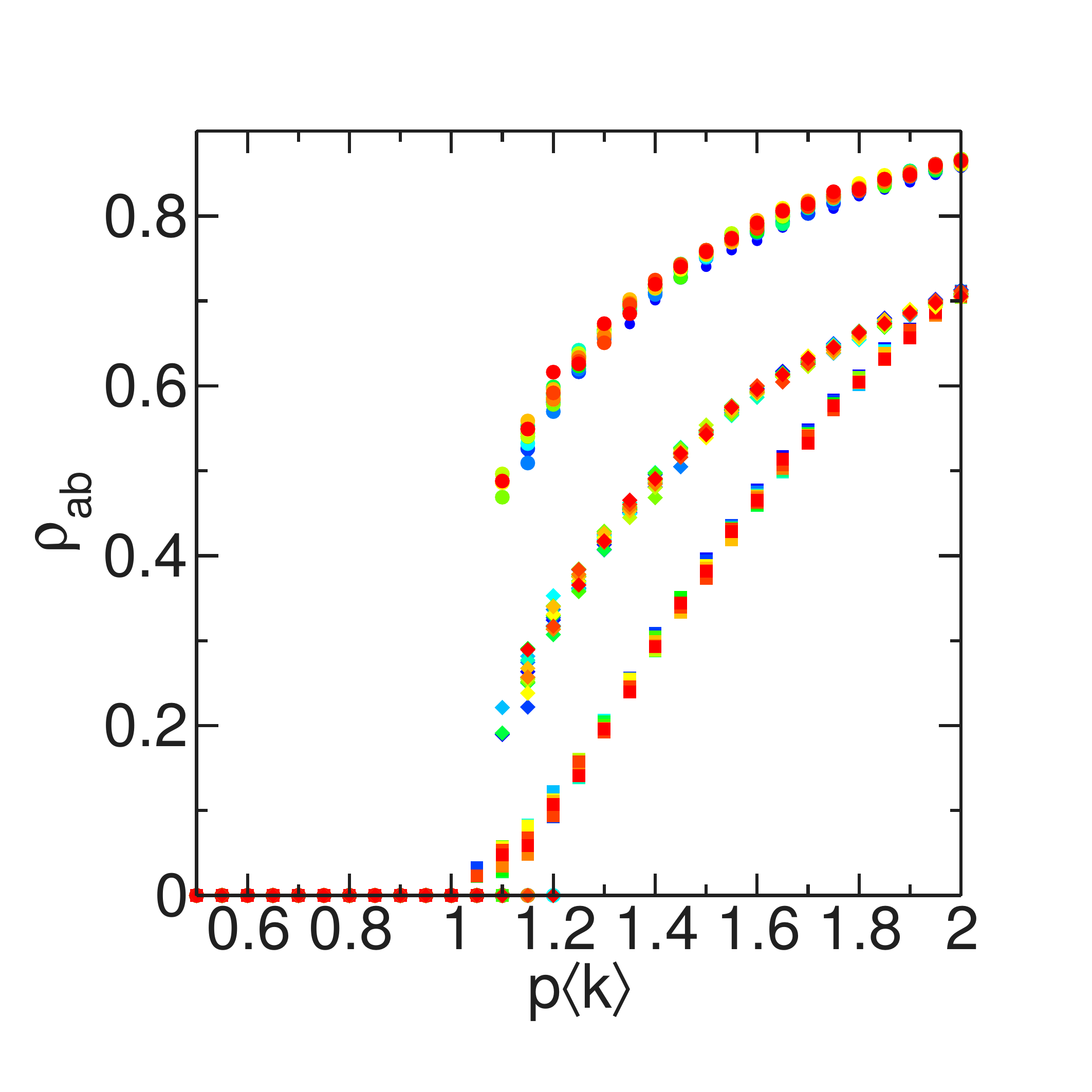}
\caption{Cooperative contagion with interactions defined by uncorrelated contact sequences. The system has a phase transition at $p_c\langle k \rangle=1$, which becomes discontinuous as the secondary infection probability $q$ is increased. Symbols denote different values of $q$: $q=p$ (squares), $q\langle k \rangle=2$ (diamonds) and $q\langle k \rangle=3$ (circles). Colors represent different interaction radii $d$, changing gradually from $d=20$ (blue) to $d=100$ (red) with $\Delta d=5$.}
\label{infvcollapse}
\end{figure}

In this section, we consider the contagion of two cooperative infections which, unless otherwise stated, have a maximum cooperative interaction ($q=1$).

Focusing on the static particles case ($v=0$), there are no finite outbreaks for $d<d_c$. For connected systems with short-range interactions ($d\gtrsim d_c$), the epidemic threshold $p_c$ is non-linear with $\langle k \rangle^{-1}$ (Fig. S5). While for $d\gtrsim d_c$ the transition is continuous (Fig. S6), long-range interactions appear as $d$ is increased, facilitating the presence of discontinuous transitions \cite{grassberger2016phase}, with the critical point shifting from $p_c\langle k \rangle=1$ to $p_c\langle k \rangle=0.5$ (Fig. S5). However, in the case $d\gg d_c$, $p_c$ has a linear growth with the inverse average degree \cite{estrada2016epidemic} $\langle k \rangle^{-1}$ , and the transition is continuous (Fig. S7), due to the immediate secondary infection of the particles after a primary infection ($q\gg p_c$); hence, the dynamics is driven by primary infections, which lead to continuous transitions. However, if we decrease $q$ for high $d\gg d_c$, $p_c$ grows, shifting from 0.5 to 1, and having a discontinuous transition (Figs. S8, S9).

In the uncorrelated contact sequence limit ($v\rightarrow \infty$), we find a universal shape for all $d$, when $p$ and $q$ are normalized with $\langle k \rangle^{-1}$ (Fig. \ref{infvcollapse}). For non-interacting infections ($q=p$), there is a continuous transition at $p\langle k \rangle=1$. However, as $q\langle k \rangle$ is increased, a discontinuous transition appears, with a characteristic gap that grows for higher values of $q$ (Figs. S10, S11). This behavior is qualitatively similar to that already reported for Erd\"os-R\'enyi networks \cite{cai2015avalanche}, with the same $p_c$ for low $q\langle k \rangle$, but leading to a change in the nature of the transition when $q$ is varied. In fact, for $q=1$, the behavior displayed by these annealed networks is understood taking into account the results on mean-field (Fig. S11) \cite{chen2013outbreaks}. Specifically, for $d_{\text{min}}<d<d_c$, the system displays a continuous transition at $p_c \langle k \rangle=1$; in fact, $q=1$ is not higher enough than $p_c$ for observing a discontinuous transition. As $d$ is increased ($d \gtrsim d_c$), a discontinuous transition appears (Fig. S12) and $p_c \langle k \rangle$ shifts towards lower values. Finally, for $d\gg d_c$, $p_c\langle k \rangle\approx 0.5$, exhibiting a continuous transition, in agreement with the results found for the static case with $d\gg d_c$, as the random redistribution of the particles, with $\langle k \rangle \sim N$, does not imply large changes on the topology.

\subsubsection*{Finite particle velocity}

\begin{figure*}[bt!]
\centering
\includegraphics[width=\textwidth]{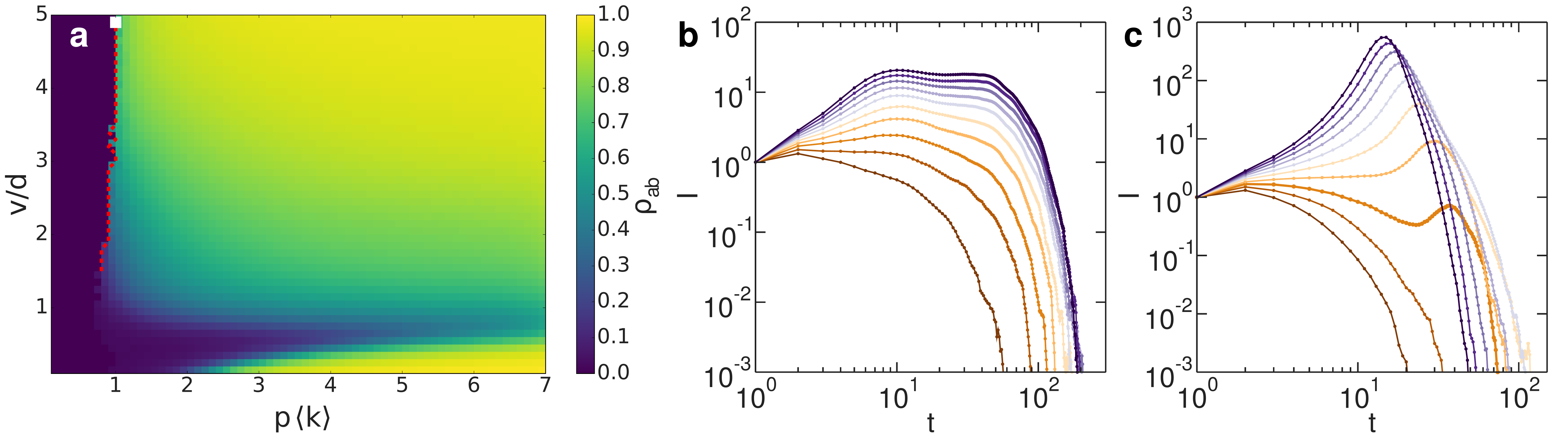}
\caption{Cooperative contagion on mobile particles with velocity $v$ for $d=30\gtrsim d_c$ and $q=1$. {\bf a}, The fraction $\rho_{ab}$ of doubly recovered particles in the final absorbing configuration has a first order phase transition for high velocities $v$. The dashed red line indicates the discontinuous transition point, while the square symbol represents $p\langle k \rangle =1$. The nature of the transition is confirmed by the time evolution of the number of infected particles, which is {\bf b}, continuous for $v=0.9d$ and {\bf c}, discontinuous for $v=5d$. The plotted curves correspond to $p\langle k \rangle= 0.7$ (brown), 1.6 (dark violet), with $\Delta (p\langle k \rangle)=0.1$.}
\label{finitev}
\end{figure*}

We vary $v$ to describe how the differences between the limits of static particles and the uncorrelated contact sequence arise. We focus on the interaction range $d=30\gtrsim d_c$ with $q=1$, which has a discontinuous transition for $v\rightarrow\infty$ (Fig. S12), while the static case lead to a continuous transition (Fig. S6), and corresponds to the regime associated with the non-monotonic behavior of $\rho$ with $v$ for a single infection (Fig. \ref{onedisvaryv}b). 

There is a continuous transition for low velocities, with $p_c$ following a non-monotonic behavior with $v$ (Fig. \ref{finitev}a,b). However, in the region associated with global mixing effects, the transition point $p_c$ does not vary too much, but the lines of constant $\rho_{ab}$ are shifting to lower values of $p$ as $v$ is increased, leading to the appearance of a discontinuous transition with a gap which grows with the velocity (Figs. \ref{finitev}a,c). 

The appearance of a discontinuous transition is explained taking into account previously reported mechanisms for this dynamics \cite{cai2015avalanche}. The gap in $\rho_{ab}$ at $p_c$ represents the occurrence or not of avalanches of secondary infections in our system. The avalanches occur when the two infections $\mathcal{A}$ and $\mathcal{B}$ meet after following different relatively long paths in the network. As $q=1$, when this meeting event occurs, an avalanche of secondary contagion events spreads over the paths that the infections followed independently previously. The requirements for this phenomenon are that the contact sequence describing the interactions has loops, such that this will not happen in static trees, and that the infections do not meet after short paths, implying that network has a low level of local clustering. Specifically, in static networks, this dynamics leads to continuous transitions in low-dimensional lattices (1D and 2D), and to discontinuous transitions for higher dimensions (4D lattices and Erd\"os-R\'enyi networks) \cite{cai2015avalanche,grassberger2016phase}. 

In our system, the mobility allows that the infections spread to a higher (and located further) fraction of particles and also decreases the effect of the local clustering, as the movement hinders the meeting events after a short path. Then, the two infections follow different and long paths before meeting: when $p=p_c$, if the two infections meet, there will be a high fraction of doubly recovered particles $\rho_{ab}$ in the final absorbing configuration; otherwise, the fraction of singly recovered particles will belong to a continuously growing branch, but the fraction of doubly recovered particles will tend to zero. Hence, there will be a discontinuous transition with two branches, with the probability of being in the upper branch representing the probability of the two infections meeting after a long path of independent contagion events.

\section*{Discussion}

The dynamics of a single infection illustrated how the epidemic threshold can be controlled with the particle velocity, leading to a non-monotonic behavior with the particle velocity. For low velocities and $d\gtrsim d_c$, the dynamics lead to higher epidemic thresholds than in the static case. Specifically, we have described the mechanism leading to a dynamical fragmentation, which has already been reported in the context of disease spreading in temporal networks \cite{rodriguez2017risk}, but also in other dynamics such as the coevolving voter model \cite{vazquez2008generic}; in contrast to the later, where the evolution of the topology depends on the dynamical configuration, our approach leads to a dynamical fragmentation even when the particles' movement is independent of their states. In the infinite velocity limit, we found a scaling relationship between the epidemic threshold and the topological parameters (for $p_c \langle k \rangle =1$, $p_c=\frac{L^2}{(N-1)\pi d^2}$). These results suggest that blood or sap velocity inside different species may have evolved in order to minimize the risk of the microcolonizations by pathogens in their environment.

For the case of two cooperative infections, not only the epidemic threshold, but also the order of the phase transition varied with the particle velocity. Previous work studying this dynamics reported that the nature of the phase transitions was influenced both by the topology and the level of cooperation \cite{cai2015avalanche,grassberger2016phase}. Specifically, intermediate levels of cooperation in mean-field approximations were leading to abrupt transitions \cite{chen2013outbreaks}. However, although a qualitatively similar behavior was observed for two-dimensional lattices with long-range interactions, the phase transitions were continuous with short-range interactions. We analyzed the interplay between this dynamics and the mobility focusing on the case $d\gtrsim d_c$. While in the case of low velocities there was a continuous transition, the global mixing effects dominated for higher velocities, making the system evolve towards discontinuous transitions. This behavior may arise from the dynamical fragmentation of the short-term loops due to the high particle velocity: in contrast to the static case, where the short loops are relatively abundant, the dynamical short-term loops do not appear in the high velocity regime. This leads to the nucleation, which happens only when the population of singly recovered ($a,b$) has grown enough such that the proximal interaction between one infected $A$ ($B$) and one recovered $b$ ($a$) is more likely, leading to a cascade of secondary infection events.

We anticipate that the mechanism leading to a non-monotonic behavior with the particle velocity, arising due to the differences between front dynamics and global mixing, may appear for different dynamics on systems of mobile particles, such as synchronization or evolutionary game theory \cite{fujiwara2011synchronization,prignano2013tuning,levis2017synchronization,meloni2009effects}. For example, the non-monotonic behavior of the synchronization with the velocity of the particles is shown for low connectivity, but it disappears when the number of interacting neighbors is increased \cite{perez2015firefly,perez2017control}. Additionally, future research will explore other movement approaches, such as the Vicsek model \cite{vicsek1995novel}, which leads to collective motion through the coupling of the movement direction between proximal particles, heterogeneity in the particle velocities, or the coupling between the particle velocity and its dynamical state.

\bibliography{rodriguezetal19}

\section*{Acknowledgments}

JPR was supported by the FPU program (MECD, Spain), and thanks Naoki Masuda for his comments. J.P.R. and V.M.E. received funding from Agencia Estatal de Investigación (AEI, Spain) and Fondo Europeo de Desarrollo
Regional through the Project SPASIMM [FIS2016- 80067-P (AEI/FEDER, UE)];
we acknowledge support from the Spanish State Research Agency through the
Mar\'ia de Maeztu Program for Units of Excellence in R\&D (MDM-2017-0711 to
the IFISC Institute). FGh acknowledges support by Deutsche Forschungsgemeinschaft under grant GH 176/1-1, within the idonate program (project 345463468), and by an EU COST action CA15109 STSM; and thanks to Alex Arenas and Yamir Moreno for their comments.

\section*{Author contributions}

J.P.R. performed the computational simulations. J.P.R., F.Gh., and V.M.E. analyzed the results and contributed equally to the manuscript. 

\section*{Code availability}

The codes used for generating the results are publicly available in:

https://github.com/jorgeprodriguezg/Particle-velocity-controls-phase-transitions-in-contagion-dynamics

\clearpage
\onecolumngrid
{\bf LIST OF SUPPLEMENTARY FIGURES}

{\bf I Temporal contact network}

Figure S1

{\bf II Single infection in an uncorrelated contact sequence}

Figure S2

{\bf III Single infection: $\bf{v=0}$ and $\bf{v=0.5d}$}

Figures S3 and S4

{\bf IV Cooperative contagion with static particles}

Figures S5, S6, S7, S8 and S9

{\bf V Cooperative contagion in an uncorrelated contact sequence}

Figures S10, S11 and S12

\renewcommand{\thefigure}{S\arabic{figure}}
\setcounter{figure}{0}
\section{Temporal contact network}
\begin{figure}[h!]
\centering
\includegraphics[width=0.6\textwidth]{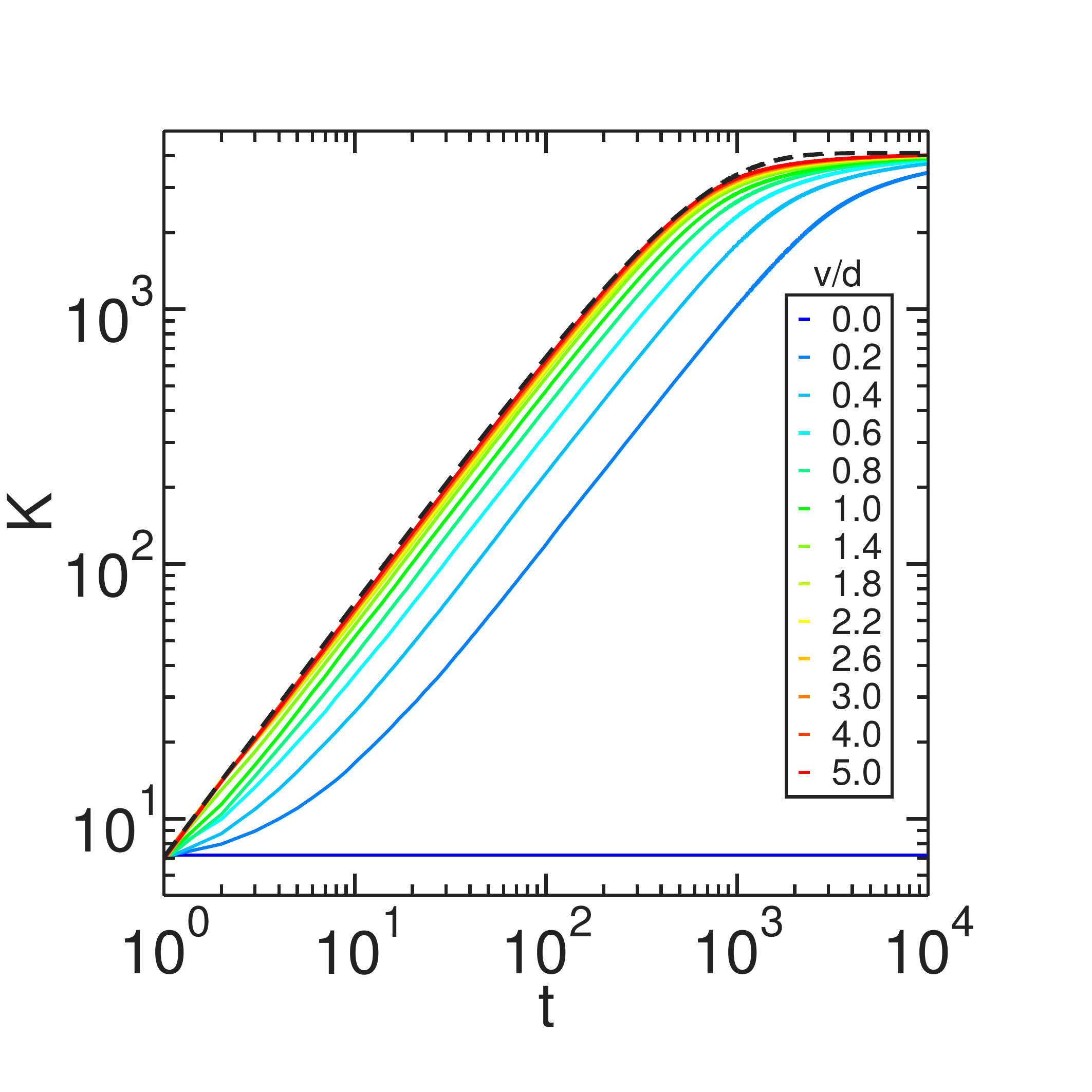}
\caption{Time evolution of the number of accumulated contacts $K$ of an average particle (\emph{i.e.}, number of particles with which it had at least one contact), for $d=30$ and different velocities, from $v=0.0$ (blue) to $v=5.0d$ (red). The black dashed line represents the uncorrelated contact network limit ($v\rightarrow \infty$), given by $K(t)=(N-1)(1-e^{-\pi d^2 t/L^2})$.}
\end{figure}
\clearpage
\section{Single infection in an uncorrelated contact sequence}

\begin{figure}[h!]
\centering
\includegraphics[width=0.6\textwidth]{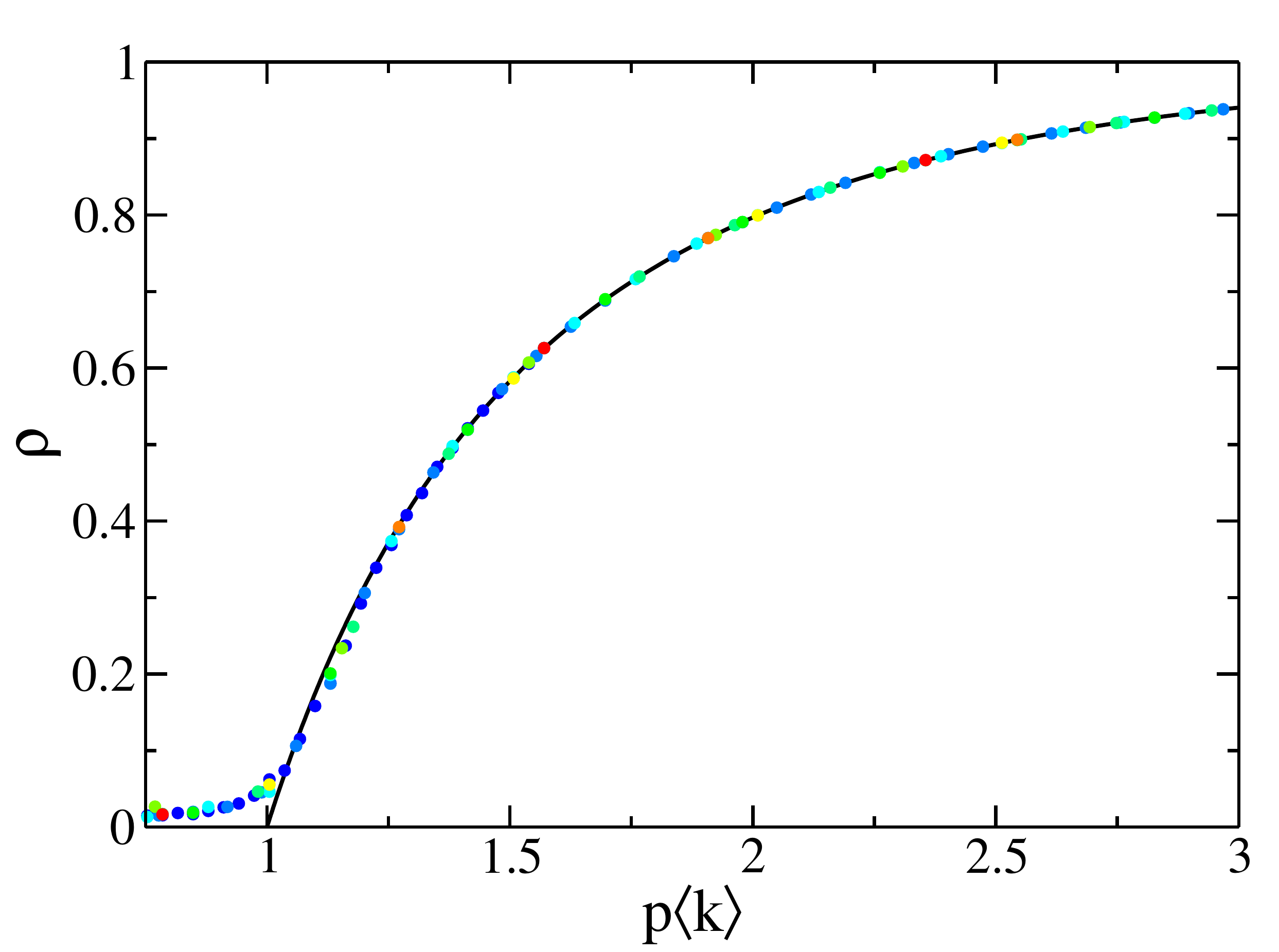}
\caption{Universal behavior of the density of recovered particles $\rho$ in the final absorbing configuration, for different interaction radii $d$ (from $d=20$ (blue) to $d=100$ (red), with colors varying continuously between these limits with $\Delta d=10$), when the control parameter is rescaled to $p\langle k \rangle$. The solid line represents the mean-field solution, given by $\rho=1-\text{e}^{p\langle k \rangle \rho}$.}
\end{figure}

\clearpage
\section{Single infection: $\bf{v=0}$ and $\bf{v=0.5d}$}
\begin{figure}[h!]
\centering
\includegraphics[width=0.8\textwidth]{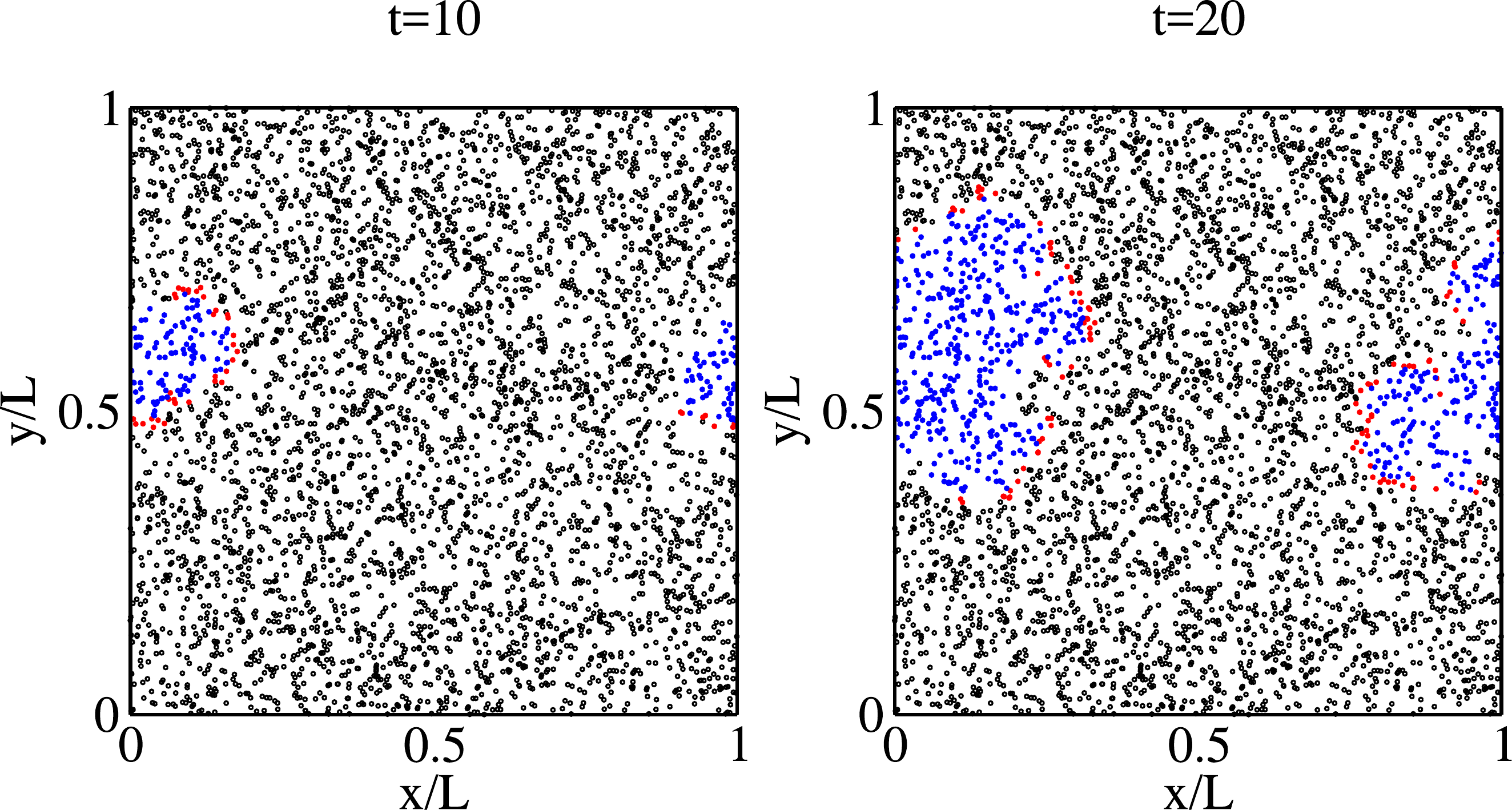}
\caption{Visualization of the contagion process at $t=10$ and $t=20$ with $p=1$, $d=30$ and $v=0$. Recovered particles (blue symbols) are connected only to whether other recovered or infected particles (red symbols), and not with susceptible particles (empty symbols).}
\end{figure}

\begin{figure}[h!]
\centering
\includegraphics[width=0.8\textwidth]{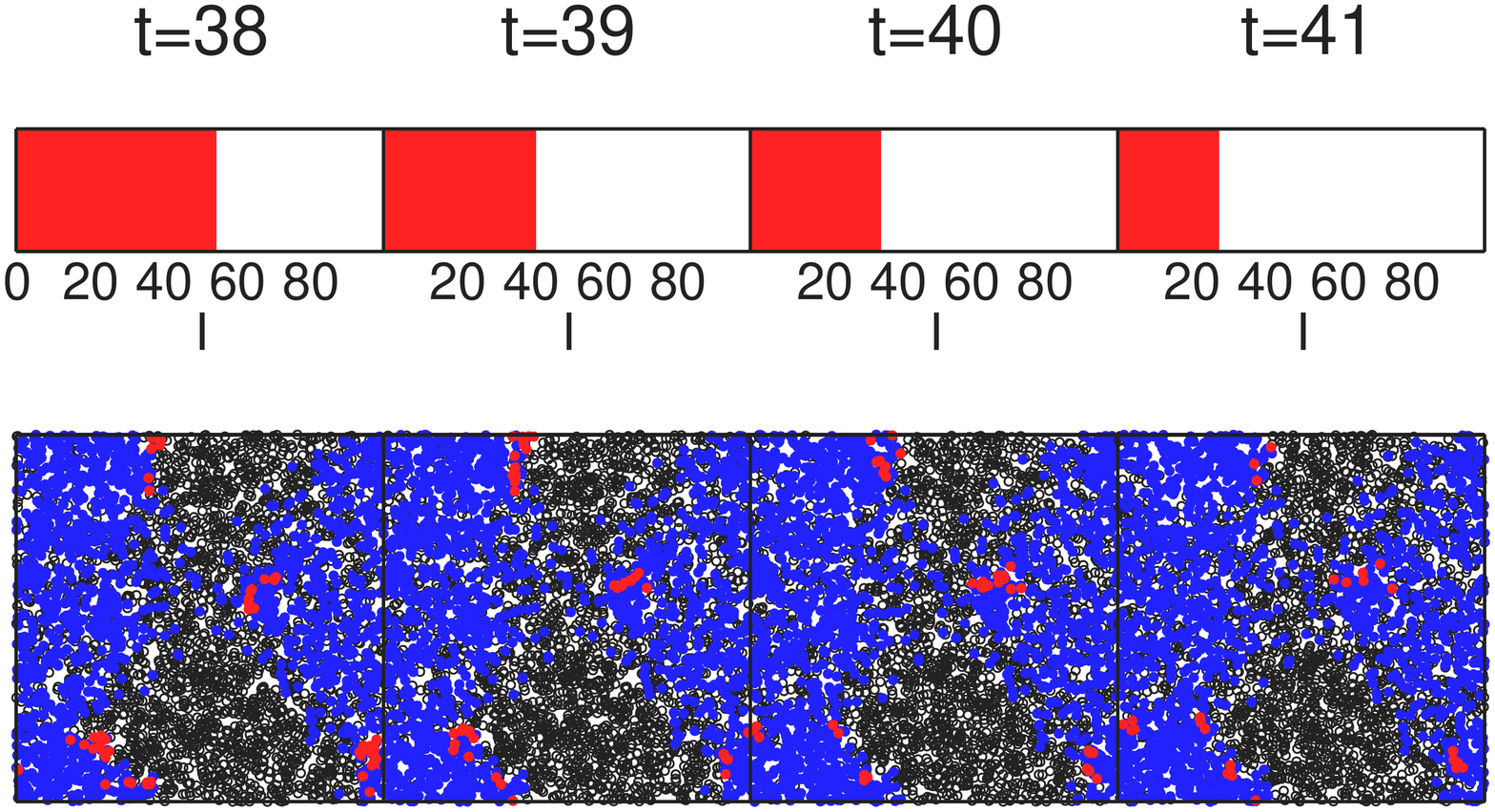}
\caption{Visualization of the contagion process at $t=38$, $t=39$, $t=40$ and $t=41$ with $p=1$, $d=28$ and $v=0.5d$. Infected particles (red symbols) are more likely connected to recovered particles (blue symbols), than to susceptible particles (empty symbols). The red upper bars represent the number of infected particles $I$.}
\end{figure}
\clearpage
\section{Cooperative contagion with static particles}
\begin{figure}[h!]
\centering
\includegraphics[width=0.6\textwidth]{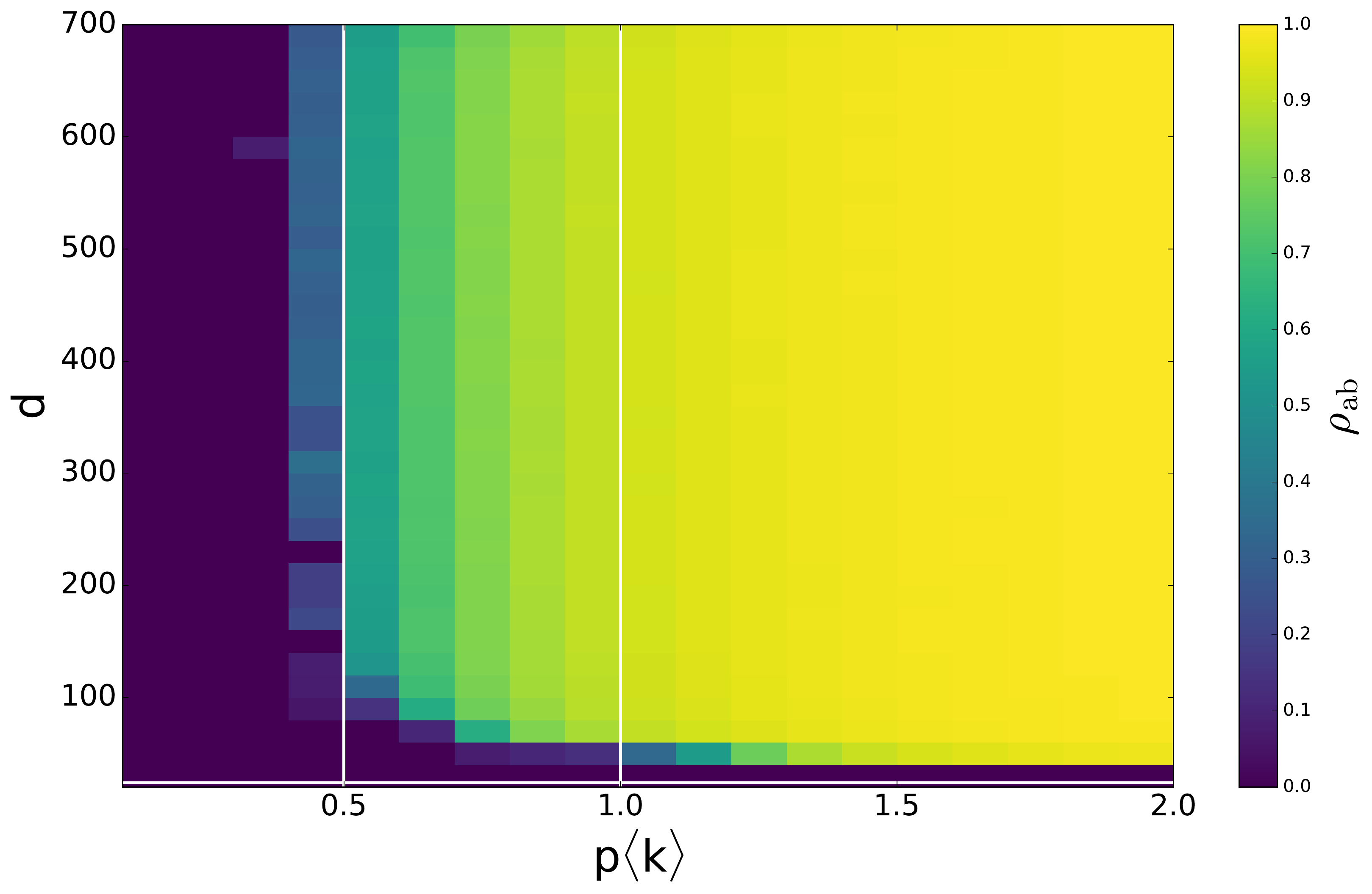}
\caption{Density of doubly recovered particles $\rho_{ab}$ in the final absorbing configuration as a function of the normalized primary infection probability $p\langle k \rangle$ and the interaction radius $d$, for $q=1$. The white horizontal line represents $d=d_c$ and the vertical lines indicate the cases $p\langle k \rangle=0.5$ and $p\langle k \rangle=1$.}
\end{figure}

\begin{figure}[h!]
\centering
\includegraphics[width=0.6\textwidth,angle=-90]{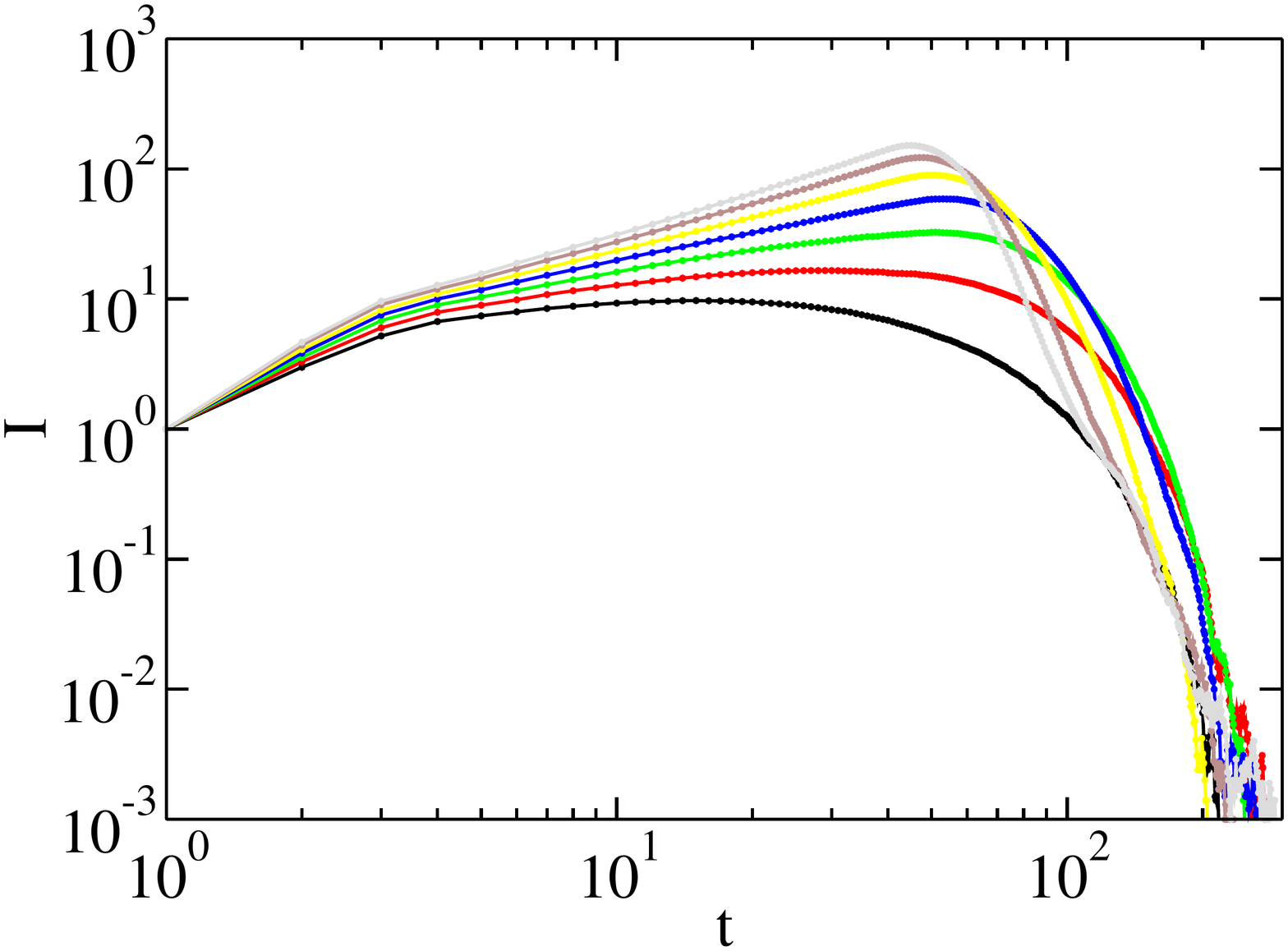}
\caption{Time evolution of the number of infected particles $I$ for a system of static particles with $d=30$ and $q=1$. The plotted values of $p\langle k \rangle$ are, from the black to the grey curve, of 1.7, 1.9, 2.1, 2.3, 2.5, 2.7 and 2.9.}
\end{figure}

\begin{figure}[h!]
\centering
\includegraphics[width=0.6\textwidth,angle=-90]{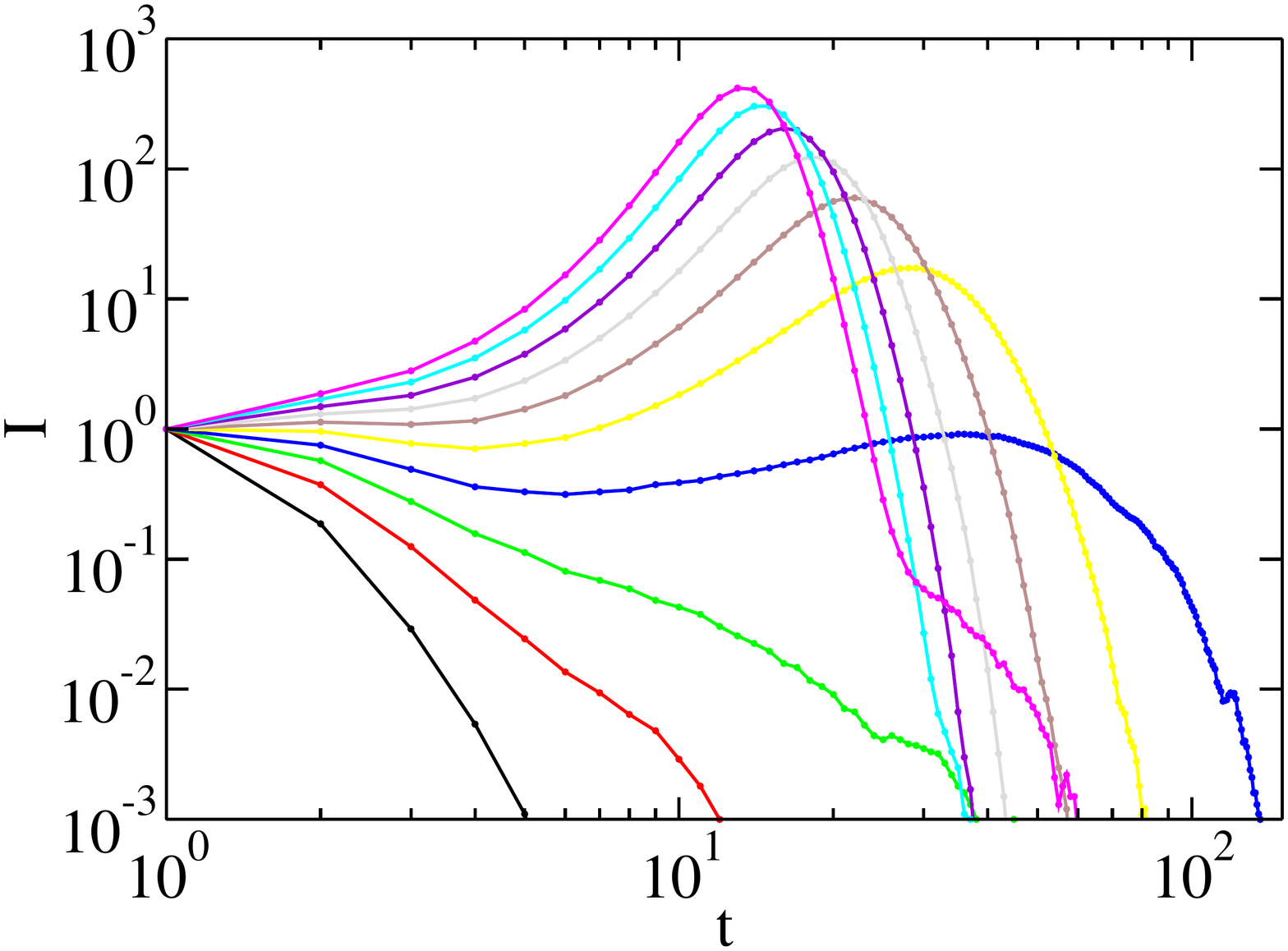}
\caption{Time evolution of the number of infected particles $I$ for a system of static particles with $d=700$ and $q=1$. The plotted values of $p\langle k \rangle$ are, from the black to the magenta curve, of 0.5, 0.6, 0.7, 0.8, 0.9, 1.0, 1.1, 1.2, 1.3 and 1.4.}
\end{figure}

\begin{figure}[h!]
\centering
\includegraphics[width=0.6\textwidth]{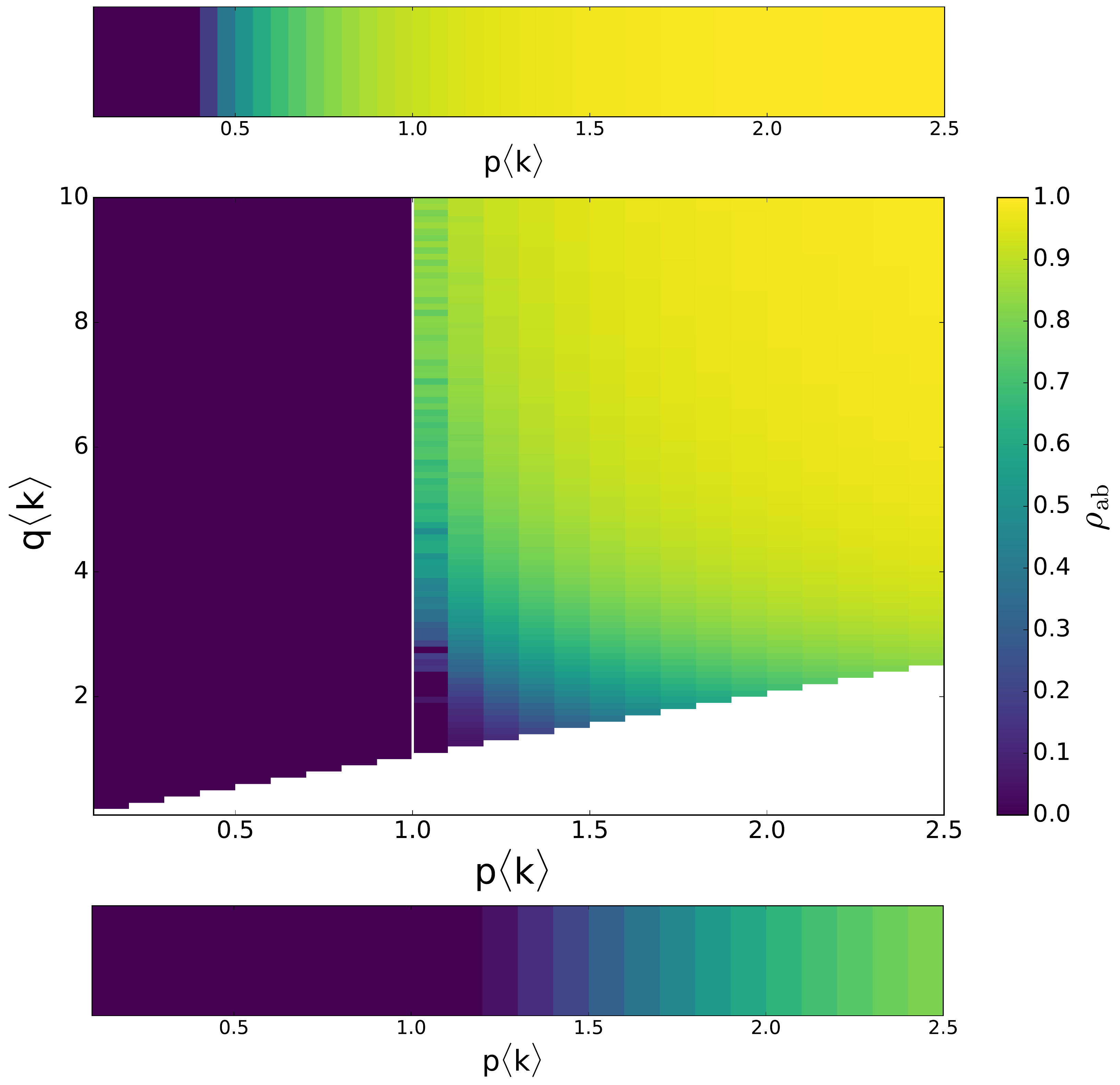}
\caption{Density of doubly recovered particles in the final absorbing configuration, as a function of $p$ and $q$, for $d=700 \gg d_c$. The upper and lower plots represent, respectively, the limit cases $q=1$ and $q=p$, and the white line in the central plot indicates the limit $p\langle k \rangle=1$.}
\end{figure}

\begin{figure}[h!]
\centering
\includegraphics[width=0.6\textwidth,angle=-90]{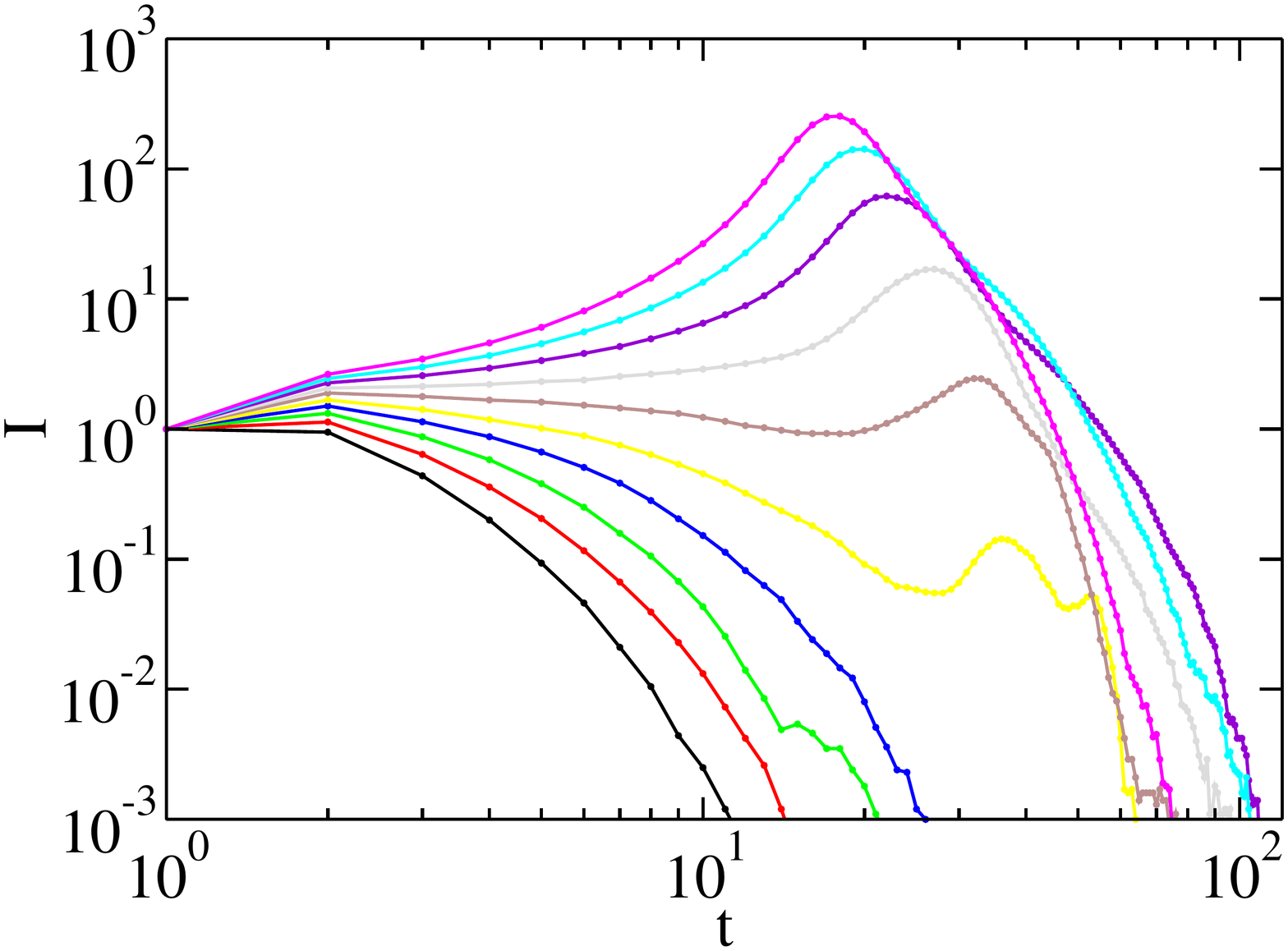}
\caption{Time evolution of the number of infected particles for a system of static particles with $d=700$ and $q\langle k \rangle=10$. The plotted values of $p\langle k \rangle$ are, from the black to the magenta curve, of 0.1, 0.2, 0.3, 0.4, 0.5, 0.6, 0.7, 0.8, 0.9 and 1.0.}
\end{figure}

\clearpage
\section{Cooperative contagion in an uncorrelated contact sequence}

\begin{figure}[h!]
\centering
\includegraphics[width=0.6\textwidth,angle=-90]{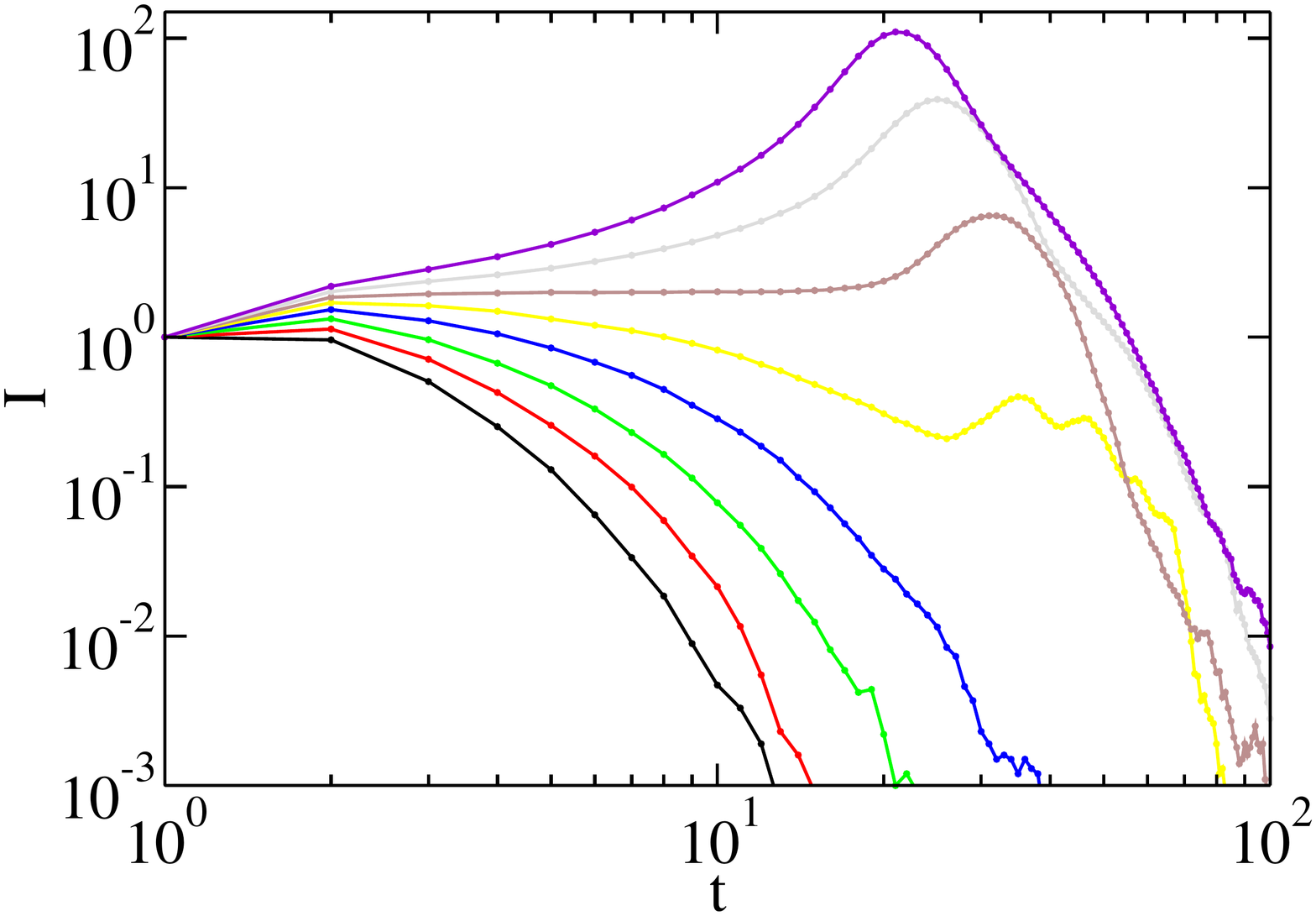}
\caption{Time evolution of the number of infected particles for $d=30$, infinite velocity and $q\langle k \rangle=3$. The plotted values of $p\langle k \rangle$ are, from the black to the violet curve, of 0.5, 0.6, 0.7, 0.8, 0.9, 1.0, 1.1, and 1.2.}
\end{figure}

\begin{figure}[h!]
\centering
\includegraphics[width=0.6\textwidth]{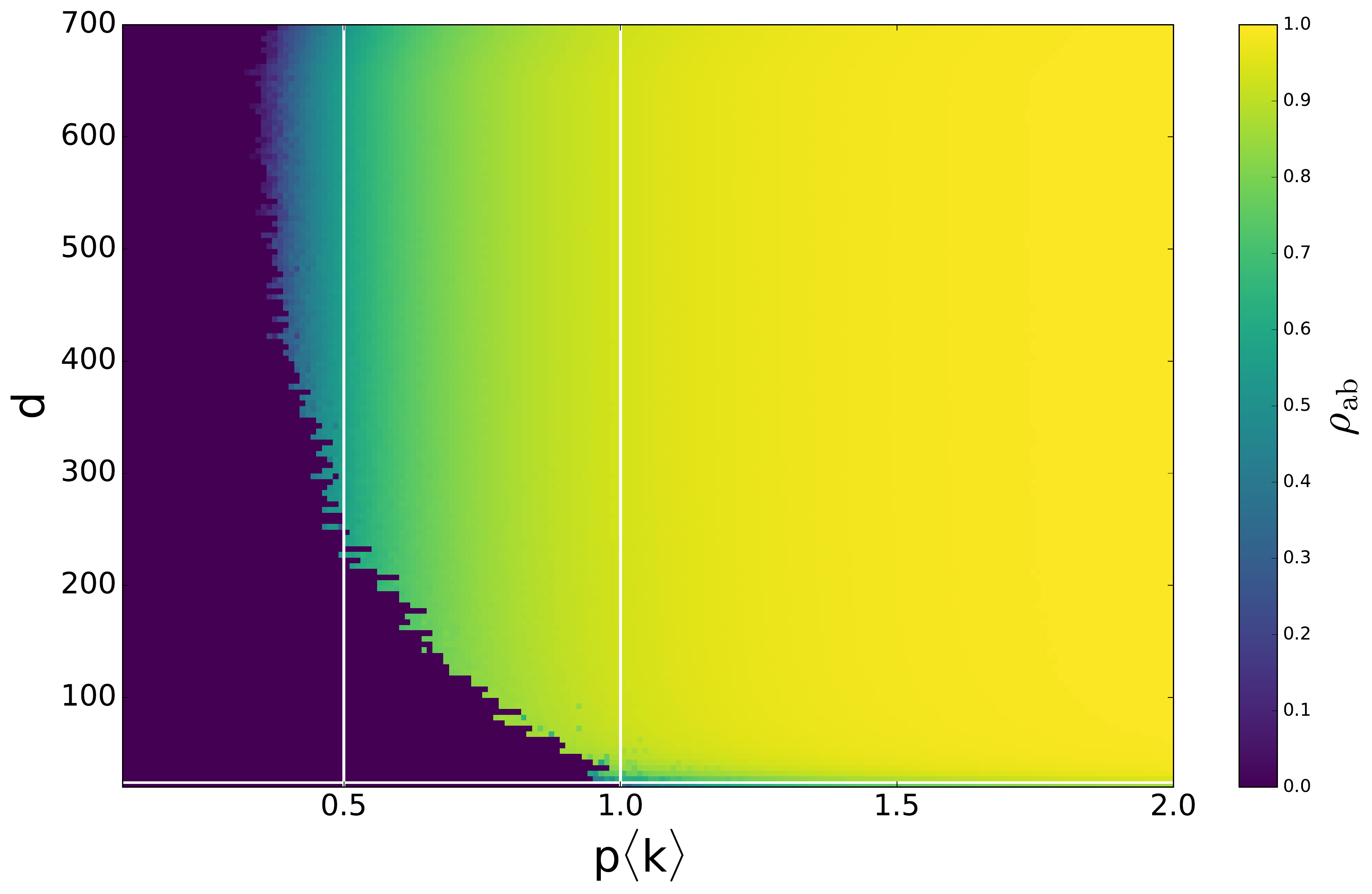}
\caption{Cooperative contagion on systems of particles with infinite velocity and $q=1$. As $d$ is increased ($d_c< d \ll d_{\text{max}}$), a continuous phase transition becomes discontinuous, coming back to a continuous phase transition for high interaction ranges, $d\sim L$. The white horizontal line represents $d=d_c$ and the vertical lines indicate the cases $p\langle k \rangle=0.5$ and $p\langle k \rangle=1$.}
\end{figure}

\begin{figure}[h!]
\centering
\includegraphics[width=0.6\textwidth]{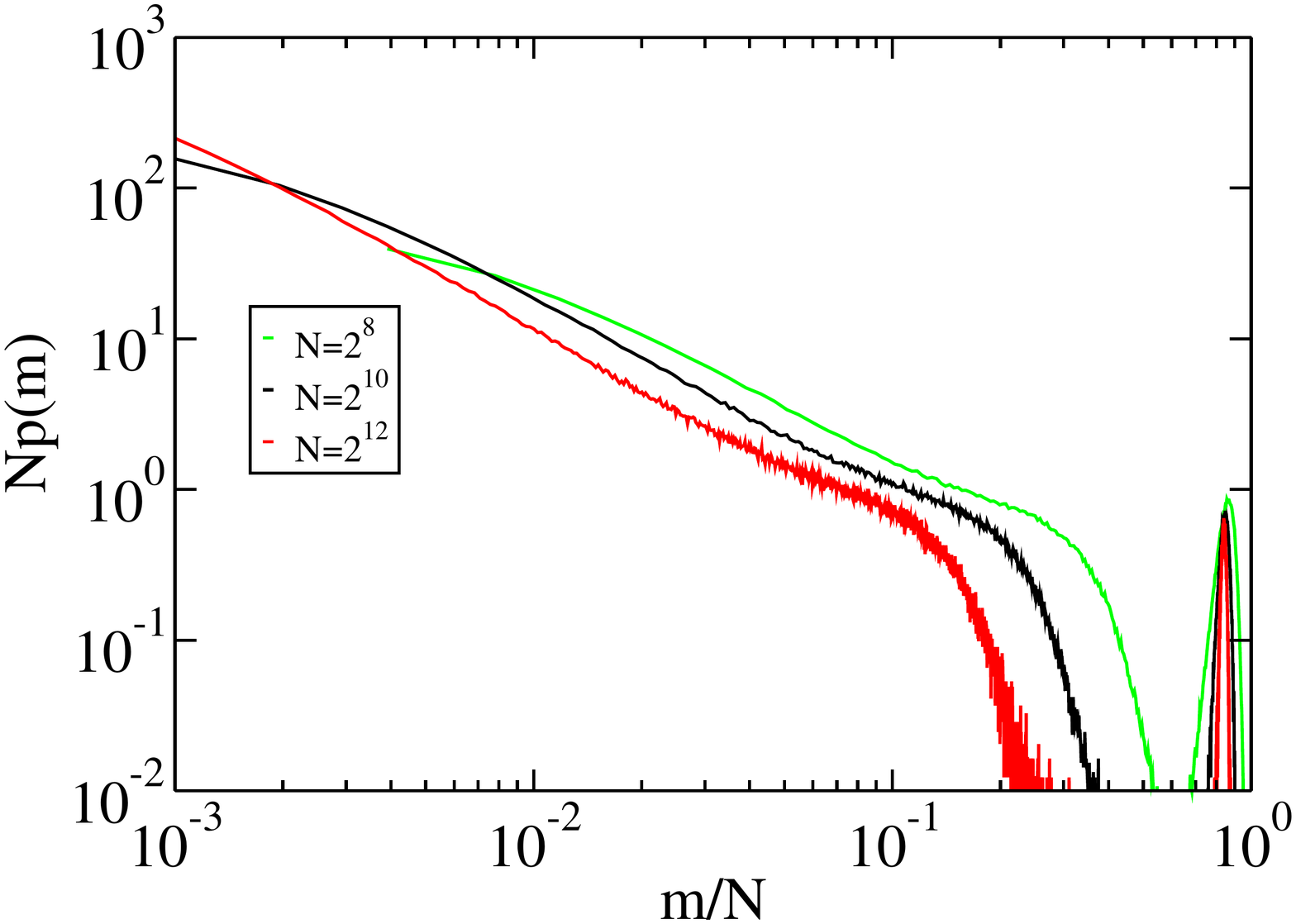}
\caption{Distribution of the number of recovered particles $m=1-S$ in the final absorbing configuration, for $d=30$, infinite velocity, $p\langle k \rangle=1$, and $q=1$, and different system sizes $N$, keeping the density $\sigma=N/L^2$ constant. There is a gap between the low prevalence configuration, leading to a fraction of recovered nodes which decays as a power-law, and the high prevalence configuration, which leads to a fraction of recovered nodes which scales linearly with the system size.}
\end{figure}

\end{document}